\documentclass{Interspeech}
\usepackage{multirow}
\usepackage{mathrsfs}

\interspeechcameraready

\title{DiffDSR: Dysarthric Speech Reconstruction Using Latent Diffusion Model}

\author{Xueyuan Chen$^1$, Dongchao Yang$^1$, Wenxuan Wu$^1$, Minglin Wu$^1$, Jing Xu$^1$, Xixin Wu$^{1,3,*}$\\
Zhiyong Wu$^{1,2,*}$, Helen Meng$^{1,2,3}$}

\affiliation{The Chinese University of Hong Kong} {Hong Kong SAR}{China}
\affiliation{Shenzhen International Graduate School, Tsinghua University}{Shenzhen}{China}
\affiliation{Vocal Engineering Technologies Limited}{Hong Kong SAR}{China}
\email{\{xychen, wuxx, hmmeng\}@se.cuhk.edu.hk zywu@sz.tsinghua.edu.cn
\thanks{* Corresponding authors. This work is supported by National Natural Science Foundation of China (62076144), CUHK Direct Grant for Research (Ref. No. 4055221), the CUHK Stanley Ho Big Data Decision Analytics Research Centre, and the Centre for Perceptual and Interactive Intelligence (CPII) Ltd., a CUHK-led InnoCentre under the InnoHK initiative of the Innovation and Technology Commission of the Hong Kong Special Administrative Region Government.}
}

\usepackage{comment}

\begin{document}

\maketitle

\begin{abstract}
    Dysarthric speech reconstruction (DSR) aims to convert dysarthric speech into comprehensible speech while maintaining the speaker's identity. 
    Despite significant advancements, existing methods often struggle with low speech intelligibility and poor speaker similarity.
    In this study, we introduce a novel diffusion-based DSR system that leverages a latent diffusion model to enhance the quality of speech reconstruction. 
    Our model comprises:
    (i) a speech content encoder for phoneme embedding restoration via pre-trained self-supervised learning (SSL) speech foundation models;
    (ii) a speaker identity encoder for speaker-aware identity preservation by in-context learning mechanism;
    (iii) a diffusion-based speech generator to reconstruct the speech based on the restored phoneme embedding and preserved speaker identity.
    Through evaluations 
    on the widely-used UASpeech corpus, our proposed model shows notable enhancements in speech intelligibility and speaker similarity.\footnote[1]
    {\href{https://Chenxuey20.github.io/DiffDSR}{Audio samples: https://Chenxuey20.github.io/DiffDSR/}}
\end{abstract}

\begin{figure*}[ht]
\centering
\includegraphics[width=2.1 \columnwidth]{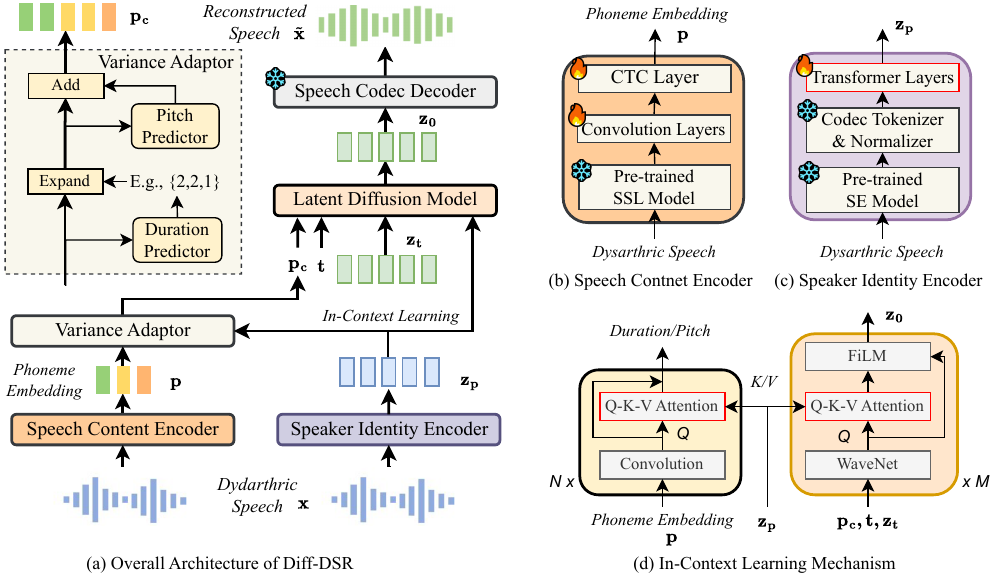}
\caption{Diagram of the proposed Diff-DSR system, where (a) shows the overall architecture, (b), (c) and (d) show the model details of speech content encoder, speaker identity encoder and in-context learning mechanism respectively.}
\vspace{-10pt}
\label{fig:model_details}
\end{figure*}

\section{Introduction}
Dysarthria is a motor speech disorder characterized by difficulty in articulating words due to weak, paralyzed, or uncoordinated muscles used for speech
\cite{ziegler2003speech}.
It can result in a notable decline in speech quality and voice characteristics compared to typical speech patterns 
\cite{sapir2014multiple}
and significantly impedes communication for dysarthria patients
\cite{freed2023motor}.
Dysarthric speech reconstruction (DSR) 
stands out as a highly effective method aimed at enhancing speech intelligibility and preserving speaker identity by 
converting dysarthric speech into a more typical speech pattern.

In comparison to the voice conversion (VC) task, DSR is a more intricate process that has attracted considerable research focus.
For individuals with less severe dysarthria in the early stages, voice banking-based methods \cite{yamagishi2012speech} aim to gather pre-recorded normal speeches before their speech abilities decline, enabling the development of personalized text-to-speech (TTS) systems \cite{chen2022hilvoice}. 
Some studies have explored adapting VC techniques to reconstruct dysarthric speech signals, such as rule-based VC \cite{kumar2016improving} and statistical VC approaches \cite{aihara2017phoneme}.
With the advancements in deep learning, 
end-to-end DSR (E2E-DSR) \cite{wang2020end} has been introduced.
It utilizes a speech encoder derived from a pre-trained automatic speech recognition (ASR) model to replace the text encoder in a sequence-to-sequence TTS system,
showing more resilient generation results than cascaded systems (ASR followed by TTS).
Besides, in order to further enhance prosody and speaker similarity, additional components such as prosody corrector and speaker encoder, have been integrated \cite{wang2020learning} 
by the similar manner 
in some TTS systems \cite{chen2022unsupervised,chen2024stylespeech}.
Additionally,
Unit-DSR \cite{wang2024unit} is proposed to use the discrete speech units extracted from HuBERT \cite{hsu2021hubert} for generating a normal speech waveform,
which aims to address the issue of training inefficiencies of complex pipelines.
To enhance speech intelligibility for severe dysarthria patients 
in noisy acoustic environments,
a multi-modal framework \cite{chen2024exploiting} has been introduced.
The visual information, e.g., lip movements, is utilized as additional clues for reconstructing the highly abnormal pronunciations,
which has been widely used in some audio-visual tasks \cite{wu2024target}.
Building upon this foundation,
CoLM-DSR \cite{chen2024colm} is further proposed by combining an audio-visual encoder with the neural codec language model framework to improve the speaker similarity and prosody naturalness.

Though significant progress has been made,
there are still two primary aspects that need to be further considered:
(i) How to effectively adapt to new speakers with more preserved speaker identities;
(ii) How to properly reconstruct the content representations 
while maintaining speaker identities.
Addressing these aspects is crucial for ensuring that the reconstructed speech aligns closely with the pronunciation habits of patients, which is essential for their sense of self-identity and guidance in rehabilitation training. 
With the advent of zero-shot 
TTS
techniques, 
language model (LM) and diffusion model are two 
potential directions.
LM-based methods \cite{wang2023neural, chen2024vall} utilize a next-token prediction manner, 
emphasizing the 
modality alignment between text and speech,
which can be challenging for severe dysarthria individuals.
In contrast, diffusion-based methods \cite{shen2023naturalspeech, ju2024naturalspeech} employ a non-autoregressive diffusion-denoising strategy, 
relying solely on the speech modality prompt, and have shown more robust results.

Inspired by the success of diffusion model in tasks such as speech generation \cite{yang2024simplespeech}, image generation \cite{peebles2023scalable} and video generation  \cite{brooks2024video}, 
this paper introduces a novel DSR system that leverages the latent diffusion model to improve the reconstruction results for the speech intelligibility and speaker similarity.
Firstly, a speech content encoder with different self-supervised learning (SSL) speech foundation models is designed to extract the robust phoneme embedding from dysarthric speech.
Secondly, we adopt a speaker identity encoder with in-context learning mechanism to 
preserve the 
speaker-aware representation from the dysarthric speech.
Finally, we use the speech generator with the latent diffusion model to reconstruct the speech based on the restored phoneme embedding condition and the preserved speaker-aware representation prompt.
The contributions of this paper include:
\begin{itemize}
\item We propose a diffusion-based DSR system that combines a speech content encoder, a speaker identity encoder and the latent diffusion model framework to reconstruct the dysarthric speech.

\item Three commonly used pre-trained SSL speech foundation models are investigated and compared to conduct the speech content restoration for the diffusion condition input. 

\item Both subjective and objective experimental results show that our proposed diffusion-based DSR system achieves notable improvements in terms of speech intelligibility and speaker similarity.
\end{itemize}

\vspace{-5pt}
\section{Methodology}
The overall architecture of our proposed Diff-DSR model is illustrated in Figure \ref{fig:model_details} (a).
It 
comprises
a speech content encoder, a speaker identity encoder and a latent diffusion-based speech generator.
The speech content encoder strives to extract robust phoneme embedding from dysarthric speech input to provide content condition.
The speaker identity encoder is designed to derive speaker-aware representation from dysarthric speech input to provide speaker identity prompt.
The speech generator utilizes the speech content condition and speaker identity prompt as inputs to reconstruct the speech.

\subsection{Speech Content Encoder for Content Restoration}
\label{sec: speech content encoder}
Firstly, a speech content encoder is devised to extract the robust linguistic representation from original dysarthric speech
for the content restoration.
Following \cite{chen2024colm}, 
we utilize the phoneme probability distribution 
embedding as the content representation.
As illustrated in Figure \ref{fig:model_details} (b), the speech content encoder comprises two main parts.
Initially, we adopt the 
SSL speech foundation models that are pre-trained on a vast amount of speech data to extract the hidden semantic features from the dysarthric speech $\mathbf{x}$.
To explore the impact of different SSL models on content restoration,
we compare and investigate three commonly used SSL speech models: 
Wav2Vec 2.0 \cite{baevski2020wav2vec}, HuBERT \cite{hsu2021hubert} and WavLM \cite{chen2022wavlm}.
which have shown significant recognition performance for low-resource languages \cite{zhao2022improving}.
Subsequently, several convolution layers are employed to 
capture local dependencies in context semantic relevance,
followed by a connectionist temporal classification (CTC) layer with CTC loss to generate the phoneme embedding $\mathbf{p}$.
During training, only the parameters $\theta_{ASR}$ of the convolution layers and CTC layer are updated to avoid overfitting.
It can be described as 
$\mathbf{p}=f_{ASR}(\mathbf{SSL} (\mathbf{x});\theta_{ASR})$,
where $\mathbf{SSL}$ is pre-trained SSL model with frozen parameters.

\subsection{Speaker Identity Encoder with In-Context Learning}
We have specially crafted a speaker identity encoder to provide speaker-aware representation prompt in our system for quick speaker identity preservation.
As depicted in Figure \ref{fig:model_details} (c), the speaker identity encoder consists of three key components:

\textbf{Pre-trained SE Model:} Initially, We utilize a pre-trained speech enhancement (SE) model, i.e., ClearerVoice \footnote[1]{\href{https://github.com/modelscope/ClearerVoice-Studio}{https://github.com/modelscope/ClearerVoice-Studio}},
to reduce irrelevant background noise, which seriously impacts the sound quality of reconstructed speech.

\textbf{Codec Tokenizer \& Normalizer:}
Then, we adopt a pre-trained neural audio codec model, EnCodec \cite{defossez2022high}, as our tokenizer.
For each input speech utterance $\mathbf{x}$, it can output a quantized and downsampled hidden representation $\hat{\mathbf{z}}_p$, where $\hat{\mathbf{z}}_p = Encodec_{enc}(\mathbf{x})$.
To not only preserve the speaker identity but also remove some dysarthric acoustic details,
following \cite{chen2024colm}, we employ a normalization operation to map the dysarthric codecs $\hat{\mathbf{z}}_p$ into corresponding normal codecs $\Tilde{\mathbf{z}}_p$ within a pre-prepared nomal codec set $\mathscr{Z}$.
This operation is facilitated by a speaker verification (SV) estimator \cite{wan2018generalized} with parameters $\theta_{SV}$ through the nearest L1 distance.
It can be formulated as:
 \begin{equation}
    \begin{split}
        & \hat{\mathbf{z}}_p \to \tilde{\mathbf{z}}_p = \arg\min_{\hat{\mathbf{z}}_p \in \mathscr{Z}} {\mid f_{SV}(\hat{\mathbf{z}}_p;\theta_{SV}) - f_{SV}(\tilde{\mathbf{z}}_p;\theta_{SV}) \mid}
    \end{split}
\end{equation}


\textbf{In-Context Learning Mechanism:}
To facilitate in-context learning for zero-shot speaker identity preservation,
we further utilize several Transformer layers to process the speaker codec prompt $\tilde{\mathbf{z}}_p$,
resulting in a hidden speaker-aware representation $\mathbf{z_p}$.
Additionally, a Q-K-V attention layer is inserted for both duration/pitch predictors in the variance adaptor and WaveNet layer in the latent diffusion model, as shown in Figure \ref{fig:model_details} (d).

\subsection{Diffusion-based Generator for Speech Reconstruction}
Inspired by the zero-shot TTS technique \cite{shen2023naturalspeech} and zero-shot VC method \cite{zhu2025zsvc}, 
we leverage the diffusion-based generator to reconstruct the dysarthric speech with the phoneme condition $\mathbf{p}$ and speaker-aware identity prompt $\mathbf{z}_p$.

\textbf{Variance Adaptor:}
In order to provide more acoustic-related features and 
alleviate the diffusion process complexity,
the variance adaptor is firstly employed to predict both duration and pitch from the phoneme embedding $\mathbf{p}$.
As shown in Figure \ref{fig:model_details} (d). the duration/pitch predictors share the same model structure with several convolution layers but with different model parameters.
The hidden sequence is then expanded to the frame level based on the duration and combined with pitch embedding to form the final condition information $\mathbf{p_c}$.

\textbf{Latent Diffusion Model:}
Similar to NaturalSpeech 2 \cite{shen2023naturalspeech}, the diffusion (forward) process and denoising (reverse) process are formulated as a stochastic differential equation (SDE) \cite{song2020score}.
The forward SDE transforms the latent $\mathbf{z}_0$ into Gaussian noise with standard Brownian motion $w_t$ and noise schedule $\beta_t$:
\begin{equation}
    \begin{split}
        \small
        & {\rm d} z_t = -\frac{1}{2}\beta_tz_t {\rm d}t+\sqrt{\beta_t}{\rm d}w_t, t\in[0,1]
    \end{split}
\end{equation}
While the reverse 
process
maps
the Gaussian noise back to $\mathbf{z}_0$:
\begin{equation}
    \begin{split}
        \small
        & {\rm d}z_t = -\frac{1}{2}(z_t+ \nabla \log p_t(z_t))\beta_t{\rm d} t,  t\in[0,1]
    \end{split}
\end{equation}
where $\nabla \log p_t(z_t)$ is the gradient of the log-density of noise data, and more details can be seen in \cite{shen2023naturalspeech}.
Specifically, we utilize WaveNet \cite{van2016wavenet} with parameters $\theta_{W}$ as the latent diffusion backbone,
which contains $M$ blocks. 
Each block includes a dilated convolution layer, a Q-K-V attention layer, and a FiLM layer \cite{perez2018film}.
As shown in Figure \ref{fig:model_details} (a), it takes the condition information $\mathbf{p_c}$, the time step $\mathbf{t}$, the current noisy vector $\mathbf{z}_t$ and the speaker identity prompt $\mathbf{z}_p$ as inputs, predicting the latent 
representation $\mathbf{z}_0$ obtained by the neural codec as output, formulated as $\mathbf{z}_0=f_{W}(\mathbf{p_c}, \mathbf{z}_t, \mathbf{t}, \mathbf{z}_p; \theta_{W})$.

\textbf{Speech Codec Decoder:}
Ultimately, the reconstructed speech $\tilde{\mathbf{x}}$ with restored content and preserved speaker identity is obtained by feeding the latent 
$\mathbf{z}_0$ into the pre-trained speech codec decoder \cite{defossez2022high}, 
formulated as $\tilde{\mathbf{x}}=Encodec_{de}(\mathbf{z}_0)$.

\section{Experiments}

\subsection{Datasets and Training Details}
Experiments are conducted on the UASpeech \cite{kim2008dysarthric}, LibriSpeech \cite{panayotov2015librispeech}, VCTK \cite{veaux2016superseded} and LibriTTS \cite{zen2019libritts} datasets.
Among them, the UASpeech corpus is a benchmark disordered speech corpus, which is recorded by an 8-channel microphone array
with some background noise.
It comprises recordings from 19 dysarthria speakers with 765 isolated words.
We use the VCTK corpus with 105 native speakers to train the SV estimator in codec normalizer.
We use the LibriTTS corpus containing 580 hours of normal speech from 2456 speakers to train the diffusion-based speech generator by teacher-forcing mode.
The details of training usage for all datasets are shown in Table \ref{tab:datasets}.

\vspace{-3pt}
\begin{table}[ht]
\caption{Details of training usage for all datasets.}
\vspace{-5pt}
\label{tab:datasets}
\centering
\begin{tabular}{ccccc}
\toprule
Dataets   & Type & Training Modules  \\ \hline
LibriSpeech \cite{panayotov2015librispeech}   & Normal & Speech Content Encoder    \\ 
UASpeech \cite{kim2008dysarthric}  & Dysarthric & Speech Content Encoder      \\ 
VCTK \cite{veaux2016superseded}  & Normal   & Codec Normalizer   \\ 
LibriTTS \cite{zen2019libritts}  & Normal  & Diffusion-based Generator     \\ 
\bottomrule
\end{tabular}
\end{table}

Similar to \cite{wang2020learning}, four speaker-dependent DSR systems are separately built for the four selected speakers (M12, F02, M16 and F04) with the lowest speech intelligibility.
All content encoders are firstly trained on LibriSpeech for 1M steps with batch size of 16, 
and then finetuned on the target speaker of UASpeech for 2k steps to improve phoneme prediction accuracy.
For fair comparison with different SSL 
models, we adopt 
the 
official pre-trained 
`Wav2Vec 2.0 Base',
`HuBERT Base'\footnote[1]
{\href{https://github.com/facebookresearch/fairseq}{https://github.com/facebookresearch/fairseq}}
and `WavLM Base'\footnote[2]{\href{https://github.com/microsoft/unilm/tree/master/wavlm}{https://github.com/microsoft/unilm/tree/master/wavlm}} models in our experiments,
all of which are pre-trained with LibriSpeech corpus without finetuning.
The diffusion-based generator is implemented based on NaturalSpeech2
and trained on 4 NVIDIA V100 GPUs for 300K iterations with a batch size of 4 on each GPU.

\subsection{Speech Intelligibility Comparison}

To comprehensively investigate the performance of different SSL models for content restoration,
we conduct the following four system settings with different content encoders for 
both objective content encoder comparison and subjective human listening test comparison.

\begin{itemize}
\item \textbf{SV-DSR (VGG)}: It uses a VGG-based  content encoder to extract phoneme embedding, and a speaker verification-based speaker encoder to represent the speaker embedding for speech reconstruction
\cite{wang2020learning}. 

\item \textbf{Diff-DSR (Wav2Vec)}: The Wav2Vec 2.0 is used 
in the speech content encoder for our proposed Diff-DSR model.

\item \textbf{Diff-DSR (HuBERT)}: The HuBERT is used in the speech content encoder for our proposed Diff-DSR model.

\item \textbf{Diff-DSR (WavLM)}: The WavLM is used in the speech content encoder for our proposed Diff-DSR model.

\end{itemize}


First of all, we conduct the content encoder comparison to verify the effectiveness of speech content encoder with different SSL models.
We decode the phoneme sequences from the phoneme embedding outputs from different system settings, respectively. 
Phoneme error rate (PER) is calculated with the ground-truth phoneme sequences. 
The results are shown in Table \ref{tab:objective speech intelligibility}.
We can observe that all the pre-trained SSL models
exhibit substantial enhancements in phoneme prediction performance when compared to the VGG-based baseline.
Moreover, the system utilizing the WavLM model demonstrates superior results across all patients in comparison to the Wav2Vec 2.0 and HuBERT models.
These results underscore the robust capability of WavLM for this low-resource recognition task, 
which are consistent with some previous studies for the similar low-source language recognition tasks \cite{zhao2022improving}.

\vspace{-3pt}
\begin{table}[ht]
\caption{Content encoder comparison results of phoneme error rate (PER) for speech intelligibility.}
\vspace{-5pt}
\label{tab:objective speech intelligibility}
\centering
\begin{tabular}{ccccc}
\toprule
Models   & M12  & F02  & M16  & F04 \\ \hline 
SV-DSR (VGG)   & 62.1\%  & 49.1\% & 46.5\% & 43.0\%      \\ 
Diff-DSR (Wav2Vec)  & 61.8\%  & 42.4\% & 39.7\% & 35.1\%     \\ 
Diff-DSR (HuBERT)  & 61.9\%  & 41.8\% & 39.4\% & 34.6\%     \\ 
Diff-DSR (WavLM)  & \textbf{61.3\%}  & \textbf{40.3\%} & \textbf{37.1\%} & \textbf{33.4\%}     \\ 
\bottomrule
\end{tabular}
\end{table}


Considering
the open-sourced ASR tools 
are not as robust as humans and can be easily influenced by sound quality and speaking styles,
following \cite{wang2024unit}, we further conduct a human listening test (HLT) to evaluate the speech intelligibility of final reconstructed speech.
30 common words are randomly selected in the test set and 5 listeners are invited to judge whether the utterance corresponds to the word transcription.
As shown in Table \ref{tab:subjective speech intelligibility},
it is evident that as the severity of dysarthria increases, the original speech becomes more challenging for others to recognize, highlighting the impact of dysarthria on speech intelligibility.
The DSR method proves to be a relatively effective approach for enhancing speech intelligibility.
Consistent with the PER results, the Diff-DSR (WavLM) system achieves the best performance across all patients compared to the other system settings.
Both the PER and HLT results show the remarkable performance of pre-trained WavLM model for this DSR task.

\vspace{-3pt}
\begin{table}[ht]
\caption{Comparison results of human listening test (HLT) accuracy for speech intelligibility.}
\vspace{-5pt}
\label{tab:subjective speech intelligibility}
\centering
\begin{tabular}{ccccc}
\toprule
Models   & M12  & F02  & M16  & F04 \\ \hline
Original Speech   & 7.4\%  & 29.3\% & 43.0\% & 62.0\%   \\ 
SV-DSR (VGG)   & 32.6\%  & 38.7\% & 55.3\% & 75.3\%      \\ 
Diff-DSR (Wav2Vec)  & 33.3\%  & 41.3\% & 60.0\% & 77.3\%     \\ 
Diff-DSR (HuBERT)  & 34.0\%  & 42.0\% & 60.7\% & 78.0\%     \\ 
Diff-DSR (WavLM)  & \textbf{34.6\%}  & \textbf{43.3\%} & \textbf{62.0\%} & \textbf{78.7\%}     \\ 
\bottomrule
\end{tabular}
\end{table}
\vspace{-10pt}

\subsection{Speaker Similarity Comparison}

To facilitate a fair comparison of speaker similarity without being influenced by content errors,
we randomly select 10 utterances with accurately restored content for each of four dysarthric speakers.
We conduct the following 
settings for both subjective and objective speaker similarity comparisons:

\begin{itemize}
\item \textbf{SV-DSR}: It uses a speaker encoder to extract a global timbre embedding and a multi-speaker mel-based decoder \cite{wang2020learning}.
\item \textbf{CoLM-DSR}: 
Excluding the influence of multi-modal input, it uses a LM-based generator with speech codec prompt \cite{chen2024colm}. 
\item \textbf{Diff-DSR(ab)}: An ablation setting of our proposed model without the codec normalizer in speaker identity encoder. 
\item \textbf{Diff-DSR}: Our complete proposed diffusion based system.

\end{itemize}

\begin{figure}[t]
\centering
\includegraphics[width=1.02\columnwidth]{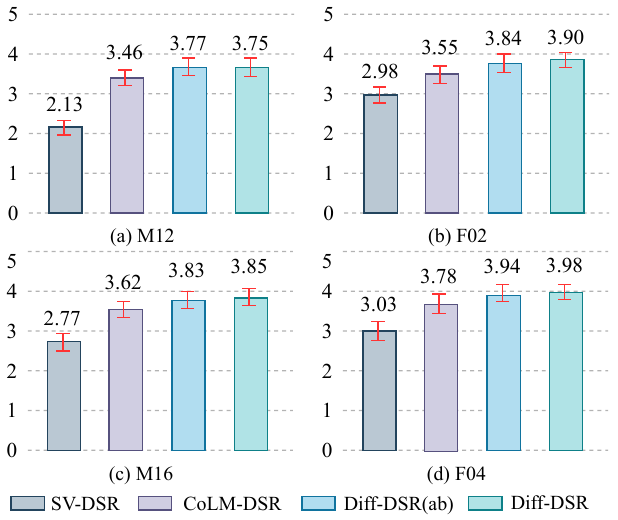}
\caption{Comparison results of MOS with 95\% confidence in terms of speaker similarity.}
\label{fig:mos}
\vspace{-10pt}
\end{figure}

We conduct the subjective test to evaluate the speaker similarity of reconstructed speech compared with the original dysarthric speech.
10 subjects are invited to give the 5-point mean opinion score (MOS, 1-bad, 2-poor, 3-fair, 4-good, 5-excellent). 
The scores are averaged and depicted in Figure \ref{fig:mos}.
As can be observed, the multi-speaker mel-based system exhibits the poorest performance for all patients.
Compared with the codec LM-based system, our proposed diffusion-based model achieves significant similarity improvements for all the 4 speakers.
What's more, the Diff-DSR shows comparable or better results with Diff-DSR(ab) in speaker similarity but significantly enhanced prosody and sound quality in practice, indicating the necessity of codec normalizer in our proposed model.

Besides, we also employ the speaker verification model \cite{wan2018generalized} as an objective measure to evaluate the speaker similarity. 
The L1 distances between the dysarthric speeches and corresponding reconstructed speeches are calculated and results are shown in Table \ref{tab:objective speaker similarity}.
Similar to the MOS findings, our proposed model also achieves the best results.
Both the subjective and objective results highlight that our proposed diffusion based DSR system can preserve more speaker information benefited from the zero-shot speaker identity preservation  ability of latent diffusion model and in-context learning mechanism.
Compared with the multi-speaker mel-based system and codec LM-based method, our diffusion-based model shows greater robustness and suitability for 
this low-source DSR task.

\vspace{-3pt}
\begin{table}[ht]
\caption{Objective comparison results for speaker similarity.}
\vspace{-5pt}
\label{tab:objective speaker similarity}
\centering
\begin{tabular}{ccccc}
\toprule
Models   & M12  & F02  & M16  & F04 \\ \hline
SV-DSR    & 1.122  & 1.151 & 1.073 & 1.056      \\ 
CoLM-DSR  & 1.081  & 1.076 & 0.985 & 0.969     \\ 
Diff-DSR(ab)  & \textbf{1.073}  & 1.071 & 0.976 & 0.958     \\ 
Diff-DSR  & 1.075  & \textbf{1.070} & \textbf{0.973} & \textbf{0.955}     \\ 
\bottomrule
\end{tabular}
\end{table}
\vspace{-10pt}

\section{Conclusion}

This paper proposes to leverage the latent diffusion model to enhance dysarthric speech reconstruction results. 
We explore and compare three widely used SSL speech foundation models for content restoration.
Additionally, we adopt a speaker identity encoder with in-context learning mechanism to facilitate speaker-aware identity preservation.
Furthermore,
we use the latent diffusion model to reconstruct the speech 
based on the content condition and speaker identity prompt.
Through a series of subjective and objective experiments conducted on the UASpeech corpus, our proposed Diff-DSR system shows notable improvements in terms of speech intelligibility and speaker similarity.


\bibliographystyle{IEEEtran}
\bibliography{mybib}

\begin{thebibliography}{10}
\providecommand{\url}[1]{#1}
\csname url@samestyle\endcsname
\providecommand{\newblock}{\relax}
\providecommand{\bibinfo}[2]{#2}
\providecommand{\BIBentrySTDinterwordspacing}{\spaceskip=0pt\relax}
\providecommand{\BIBentryALTinterwordstretchfactor}{4}
\providecommand{\BIBentryALTinterwordspacing}{\spaceskip=\fontdimen2\font plus
\BIBentryALTinterwordstretchfactor\fontdimen3\font minus \fontdimen4\font\relax}
\providecommand{\BIBforeignlanguage}[2]{{%
\expandafter\ifx\csname l@#1\endcsname\relax
\typeout{** WARNING: IEEEtran.bst: No hyphenation pattern has been}%
\typeout{** loaded for the language `#1'. Using the pattern for}%
\typeout{** the default language instead.}%
\else
\language=\csname l@#1\endcsname
\fi
#2}}
\providecommand{\BIBdecl}{\relax}
\BIBdecl

\bibitem{ziegler2003speech}
W.~Ziegler, ``Speech motor control is task-specific: Evidence from dysarthria and apraxia of speech,'' \emph{Aphasiology}, vol.~17, no.~1, pp. 3--36, 2003.

\bibitem{sapir2014multiple}
S.~Sapir, ``Multiple factors are involved in the dysarthria associated with parkinson's disease: a review with implications for clinical practice and research,'' \emph{Journal of Speech, Language, and Hearing Research}, vol.~57, no.~4, pp. 1330--1343, 2014.

\bibitem{freed2023motor}
D.~B. Freed, \emph{Motor speech disorders: diagnosis and treatment}.\hskip 1em plus 0.5em minus 0.4em\relax plural publishing, 2023.

\bibitem{yamagishi2012speech}
J.~Yamagishi, C.~Veaux, S.~King, and S.~Renals, ``Speech synthesis technologies for individuals with vocal disabilities: Voice banking and reconstruction,'' \emph{Acoustical Science and Technology}, vol.~33, no.~1, pp. 1--5, 2012.

\bibitem{chen2022hilvoice}
X.~Chen, Q.~Huang, X.~Wu, Z.~Wu, and H.~Meng, ``Hilvoice: Human-in-the-loop style selection for elder-facing speech synthesis,'' in \emph{ISCSLP}.\hskip 1em plus 0.5em minus 0.4em\relax IEEE, 2022, pp. 86--90.

\bibitem{kumar2016improving}
S.~A. Kumar and C.~S. Kumar, ``Improving the intelligibility of dysarthric speech towards enhancing the effectiveness of speech therapy,'' in \emph{2016 International Conference on Advances in Computing, Communications and Informatics (ICACCI)}.\hskip 1em plus 0.5em minus 0.4em\relax IEEE, 2016, pp. 1000--1005.

\bibitem{aihara2017phoneme}
R.~Aihara, T.~Takiguchi, and Y.~Ariki, ``Phoneme-discriminative features for dysarthric speech conversion.'' in \emph{Interspeech}, 2017, pp. 3374--3378.

\bibitem{wang2020end}
D.~Wang, J.~Yu, X.~Wu, S.~Liu, L.~Sun, X.~Liu, and H.~Meng, ``End-to-end voice conversion via cross-modal knowledge distillation for dysarthric speech reconstruction,'' in \emph{ICASSP 2020}.\hskip 1em plus 0.5em minus 0.4em\relax IEEE, 2020, pp. 7744--7748.

\bibitem{wang2020learning}
D.~Wang, S.~Liu, L.~Sun, X.~Wu, X.~Liu, and H.~Meng, ``Learning explicit prosody models and deep speaker embeddings for atypical voice conversion,'' \emph{arXiv preprint arXiv:2011.01678}, 2020.

\bibitem{chen2022unsupervised}
X.~Chen, S.~Lei, Z.~Wu, D.~Xu, W.~Zhao, and H.~Meng, ``Unsupervised multi-scale expressive speaking style modeling with hierarchical context information for audiobook speech synthesis,'' in \emph{COLING}, 2022, pp. 7193--7202.

\bibitem{chen2024stylespeech}
X.~Chen, X.~Wang, S.~Zhang, L.~He, Z.~Wu, X.~Wu, and H.~Meng, ``Stylespeech: Self-supervised style enhancing with vq-vae-based pre-training for expressive audiobook speech synthesis,'' in \emph{ICASSP 2024}.\hskip 1em plus 0.5em minus 0.4em\relax IEEE, 2024, pp. 12\,316--12\,320.

\bibitem{wang2024unit}
Y.~Wang, X.~Wu, D.~Wang, L.~Meng, and H.~Meng, ``Unit-dsr: Dysarthric speech reconstruction system using speech unit normalization,'' \emph{arXiv preprint arXiv:2401.14664}, 2024.

\bibitem{hsu2021hubert}
W.-N. Hsu, B.~Bolte, Y.-H.~H. Tsai, K.~Lakhotia, R.~Salakhutdinov, and A.~Mohamed, ``Hubert: Self-supervised speech representation learning by masked prediction of hidden units,'' \emph{IEEE/ACM transactions on audio, speech, and language processing}, vol.~29, pp. 3451--3460, 2021.

\bibitem{chen2024exploiting}
X.~Chen, Y.~Wang, X.~Wu, D.~Wang, Z.~Wu, X.~Liu, and H.~Meng, ``Exploiting audio-visual features with pretrained av-hubert for multi-modal dysarthric speech reconstruction,'' \emph{arXiv preprint arXiv:2401.17796}, 2024.

\bibitem{wu2024target}
W.~Wu, X.~Chen, X.~Wu, H.~Li, and H.~Meng, ``Target speech extraction with pre-trained av-hubert and mask-and-recover strategy,'' \emph{arXiv preprint arXiv:2403.16078}, 2024.

\bibitem{chen2024colm}
X.~Chen, D.~Yang, D.~Wang, X.~Wu, Z.~Wu, and H.~Meng, ``Colm-dsr: Leveraging neural codec language modeling for multi-modal dysarthric speech reconstruction,'' \emph{arXiv preprint arXiv:2406.08336}, 2024.

\bibitem{wang2023neural}
C.~Wang, S.~Chen, Y.~Wu, Z.~Zhang, L.~Zhou, S.~Liu, Z.~Chen, Y.~Liu, H.~Wang, J.~Li \emph{et~al.}, ``Neural codec language models are zero-shot text to speech synthesizers,'' \emph{arXiv preprint arXiv:2301.02111}, 2023.

\bibitem{chen2024vall}
S.~Chen, S.~Liu, L.~Zhou, Y.~Liu, X.~Tan, J.~Li, S.~Zhao, Y.~Qian, and F.~Wei, ``Vall-e 2: Neural codec language models are human parity zero-shot text to speech synthesizers,'' \emph{arXiv preprint arXiv:2406.05370}, 2024.

\bibitem{shen2023naturalspeech}
K.~Shen, Z.~Ju, X.~Tan, Y.~Liu, Y.~Leng, L.~He, T.~Qin, S.~Zhao, and J.~Bian, ``Naturalspeech 2: Latent diffusion models are natural and zero-shot speech and singing synthesizers,'' \emph{arXiv preprint arXiv:2304.09116}, 2023.

\bibitem{ju2024naturalspeech}
Z.~Ju, Y.~Wang, K.~Shen, X.~Tan, D.~Xin, D.~Yang, Y.~Liu, Y.~Leng, K.~Song, S.~Tang \emph{et~al.}, ``Naturalspeech 3: Zero-shot speech synthesis with factorized codec and diffusion models,'' \emph{arXiv preprint arXiv:2403.03100}, 2024.

\bibitem{yang2024simplespeech}
D.~Yang, D.~Wang, H.~Guo, X.~Chen, X.~Wu, and H.~Meng, ``Simplespeech: Towards simple and efficient text-to-speech with scalar latent transformer diffusion models,'' \emph{arXiv preprint arXiv:2406.02328}, 2024.

\bibitem{peebles2023scalable}
W.~Peebles and S.~Xie, ``Scalable diffusion models with transformers,'' in \emph{Proceedings of the IEEE/CVF International Conference on Computer Vision}, 2023, pp. 4195--4205.

\bibitem{brooks2024video}
T.~Brooks, B.~Peebles, C.~Holmes, W.~DePue, Y.~Guo, L.~Jing, D.~Schnurr, J.~Taylor, T.~Luhman, E.~Luhman \emph{et~al.}, ``Video generation models as world simulators,'' 2024.

\bibitem{baevski2020wav2vec}
A.~Baevski, Y.~Zhou, A.~Mohamed, and M.~Auli, ``wav2vec 2.0: A framework for self-supervised learning of speech representations,'' \emph{Advances in neural information processing systems}, vol.~33, pp. 12\,449--12\,460, 2020.

\bibitem{chen2022wavlm}
S.~Chen, C.~Wang, Z.~Chen, Y.~Wu, S.~Liu, Z.~Chen, J.~Li, N.~Kanda, T.~Yoshioka, X.~Xiao \emph{et~al.}, ``Wavlm: Large-scale self-supervised pre-training for full stack speech processing,'' \emph{IEEE Journal of Selected Topics in Signal Processing}, vol.~16, no.~6, pp. 1505--1518, 2022.

\bibitem{zhao2022improving}
J.~Zhao and W.-Q. Zhang, ``Improving automatic speech recognition performance for low-resource languages with self-supervised models,'' \emph{IEEE Journal of Selected Topics in Signal Processing}, vol.~16, no.~6, pp. 1227--1241, 2022.

\bibitem{defossez2022high}
A.~D{\'e}fossez, J.~Copet, G.~Synnaeve, and Y.~Adi, ``High fidelity neural audio compression,'' \emph{arXiv preprint arXiv:2210.13438}, 2022.

\bibitem{wan2018generalized}
L.~Wan, Q.~Wang, A.~Papir, and I.~L. Moreno, ``Generalized end-to-end loss for speaker verification,'' in \emph{ICASSP 2018}.\hskip 1em plus 0.5em minus 0.4em\relax IEEE, 2018, pp. 4879--4883.

\bibitem{zhu2025zsvc}
X.~Zhu, L.~He, Y.~Xiao, X.~Wang, X.~Tan, S.~Zhao, and L.~Xie, ``Zsvc: Zero-shot style voice conversion with disentangled latent diffusion models and adversarial training,'' \emph{arXiv preprint arXiv:2501.04416}, 2025.

\bibitem{song2020score}
Y.~Song, J.~Sohl-Dickstein, D.~P. Kingma, A.~Kumar, S.~Ermon, and B.~Poole, ``Score-based generative modeling through stochastic differential equations,'' \emph{arXiv preprint arXiv:2011.13456}, 2020.

\bibitem{van2016wavenet}
A.~Van Den~Oord, S.~Dieleman, H.~Zen, K.~Simonyan, O.~Vinyals, A.~Graves, N.~Kalchbrenner, A.~Senior, K.~Kavukcuoglu \emph{et~al.}, ``Wavenet: A generative model for raw audio,'' \emph{arXiv preprint arXiv:1609.03499}, vol.~12, 2016.

\bibitem{perez2018film}
E.~Perez, F.~Strub, H.~De~Vries, V.~Dumoulin, and A.~Courville, ``Film: Visual reasoning with a general conditioning layer,'' in \emph{Proceedings of the AAAI conference on artificial intelligence}, vol.~32, no.~1, 2018.

\bibitem{kim2008dysarthric}
H.~Kim, M.~Hasegawa-Johnson, A.~Perlman, J.~Gunderson, T.~S. Huang, K.~Watkin, and S.~Frame, ``Dysarthric speech database for universal access research,'' in \emph{Ninth Annual Conference of the International Speech Communication Association}, 2008.

\bibitem{panayotov2015librispeech}
V.~Panayotov, G.~Chen, D.~Povey, and S.~Khudanpur, ``Librispeech: an asr corpus based on public domain audio books,'' in \emph{2015 IEEE international conference on acoustics, speech and signal processing (ICASSP)}.\hskip 1em plus 0.5em minus 0.4em\relax IEEE, 2015, pp. 5206--5210.

\bibitem{veaux2016superseded}
C.~Veaux, J.~Yamagishi, K.~MacDonald \emph{et~al.}, ``Superseded-cstr vctk corpus: English multi-speaker corpus for cstr voice cloning toolkit,'' 2016.

\bibitem{zen2019libritts}
H.~Zen, V.~Dang, R.~Clark, Y.~Zhang, R.~J. Weiss, Y.~Jia, Z.~Chen, and Y.~Wu, ``Libritts: A corpus derived from librispeech for text-to-speech,'' \emph{Interspeech 2019}, 2019.

\end{thebibliography}

\end{document}